\documentclass[aps,showpacs,pra,superscriptaddress,]{revtex4-2}
\usepackage{amsmath}
\usepackage{amsfonts}
\usepackage{amssymb}
\usepackage{bm}
\usepackage{graphicx}

\begin{document}

\title{Quantum Dynamics and Elementary Excitations in Superfluid  
He$^{4}$ Films at Low Temperatures }

\author{Vladimir I. Kruglov}
\affiliation{Centre for Engineering Quantum Systems, School of Mathematics and Physics, The University of Queensland, Brisbane, Queensland 4072, Australia}
\author{ Houria Triki}
\affiliation{Radiation Physics Laboratory, Department of Physics, Faculty of Sciences, Badji Mokhtar University, P. O. Box 12, 23000 Annaba, Algeria}

\begin{abstract}
We have formulated a novel quantum nonlinear Schr\"{o}dinger equation describing of superfluid He$^{4}$ in films at low temperatures. It is shown that in classical limit the found nonlinear Schr\"{o}dinger equation reduces to a system of equations which are equivalent to Boussinesq equations describing the propagation of long gravity waves in incompressible fluids. This nonlinear Schr\"{o}dinger equation leads to phonon-roton dispersion relation for elementary excitations in superfluid He$^{4}$ films at low temperatures. The quartic soliton, dark soliton, cosine and elliptic periodic wave solutions are obtained analytically as weakly excited quantum waves propagating in He$^{4}$ films. We have also shown numerically that the presented nonlinear Schr\"{o}dinger equation describes the quartic and dark solitary waves in helium films. These solitary and periodic quantum waves can find numerous important practical applications.
\end{abstract}

\maketitle

\section{Introduction}

Superfluid helium films have been the focus of intense research interest due to their distinctive properties and their role in understanding quantum phenomena. Such films offer a promising platform for significant applications in areas like optomechanics \cite{R1}, neutron storage \cite{R2} and dark matter detection \cite{R3}. It is worth to point out that researches on these films have explored relevant physical phenomena like superfluidity in confined geometries such as thin films \cite{R4}, the behavior of helium films on different surfaces \cite{R5} and the interaction
of third sound with optical resonators \cite{R6}. These investigations are motivated by the potential for superfluid helium films to host a broad range of quantum phenomena, which makes them important for both practical application and theoretical research.
Especially, films of $^{4}$He have attracted a great deal of attention from both theoretical and experimental points of view as they demonstrate superfluidity at extremely low temperatures, which enables for the observation of macroscopic quantum phenomena.

In order to understand the behavior of superfluid helium films, particularly in confined geometries, theoretical models have been successfully developed. Such models show that highly confined superfluid films are extremely nonlinear mechanical resonators, offering the prospect to realize a mechanical qubit \cite{R7}. 
In this context, Rutledge et al. \cite{Rut} developed a two-dimensional quantum hydrodynamics and derived an expression for third-sound velocity at finite temperature. Ambegaokar et al. \cite{Amb} presented a theory of the dynamical properties
of a helium film near its superfluid transition. Sreekumar and Nandakumaran \cite{Sreek} showed that the dynamics of saturated two-dimensional superfluid $^{4}$He films is governed by the Kadomtsev-Petviashvili (KP) equation with negative dispersion in the small amplitude regime. They also found that soliton resonance could happen at lowest order nonlinearity, if two dimensional effects are considered.

Besides the development of theoretical models for superfluid He$^{4}$ films, many influential works have devoted to discuss the possibility of soliton propagation in such system. For instance, Huberman \cite{Hub} showed that in monolayer superfluid He$^{4}$ films, small finite amplitude effects can lead to the existence of gapless solitons made up of superfluid condensate. His argument starts from the Pitaevskii-Gross equation \cite{Pit,Gross} of the Bose condensate at zero temperature. He also showed that the dynamics of superfluid density is governed by the Korteweg-de Vries (KdV) equation. However, the derivation of KdV equation is not given in detail. The KdV equation is also derived by Nakajima et al. \cite{Sado} from Landau two-fluid hydrodynamics \cite{Land,Patt} applied to the thickness oscillation of the superfluid He$^{4}$ film at low temperatures.  They have found the KdV equation of motion for the film thickness by including appropriate nonlinear terms in the usual two-fluid hydrodynamics. In this paper the main restoring force is van der Waals attraction from the substrate and thermomechanical force due to phonons is a small correction. Biswas and Warke \cite{Biswas} investigated the stability of solitons in two-dimensional superfluid $^{4}$He films and suggested that two-dimensional \textquotedblleft lumps\textquotedblright\ of superfluid condensate might be experimentally observable. They also obtained the KP equation for the superfluid surface density fluctations and concluded that two-dimensional localized waves (lumps) that decay algebraically in all horizontal directions should be detectable at very low temperatures. Condat and Guyer \cite{Condat} discussed the possibility of propagating KdV solitons in a superfluid $^{4}$He film and in a superfluid $^{4}$He film overlayed by a $^{3}$He film. However, in all these papers the solitary waves are found analytically within the framework of nonlinear integrable equations like KdV and KP equations which need special physical situations where these models describe the dynamics of nonlinear waves.  More recently, Ancilotto et.al. \cite{Ancilotto} have demonstrated the first experimental observation of bright solitons in bulk superfluid $^{4}$He. 

In this paper, we describe the quantum solitary and periodic waves propagating in superfluid $^{4}$He films at low temperatures by the novel nonlinear Schr\"{o}dinger equation. We have found the quartic soliton, dark soliton, cosine and elliptic periodic wave solutions as weakly excited quantum waves propagating in He$^{4}$ films. We also have shown that this nonlinear Schr\"{o}dinger equation leads to phonon-roton dispersion relation for elementary excitations in superfluid He$^{4}$ films at low temperatures. We emphasize that all results obtained in this paper are based directly on the novel nonlinear Schr\"{o}dinger equation. It is shown that in classical limit the found nonlinear Schr\"{o}dinger equation leads to Boussinesq and KdV equations (see Appendix A). In the present paper, we restrict ourselves to the low-temperature range, where the evaporation rate is exponentially low and quantum dynamics in superfluid He$^{4}$ films is essential. 

The paper is structured as follows. In Sec. II, we present a new nonlinear Schr\"{o}dinger equation describing the quantum dynamics in superfluid $^{4}$He films at low temperatures. It is also shown in this Sec. II that in classical limit the found nonlinear Schr\"{o}dinger equation reduces to a system of equations which are equivalent to Boussinesq equations describing the propagation of long gravity waves in incompressible fluids. In Sec. III, we derive the generalized dispersion equation for superfluid helium films which takes into account the roton excitations in helium superfluid. In Sec. IV, we have derived the equation for weakly excited quantum waves propagating in He$^{4}$ films and in Sec. V, we have found
the exact solutions of nonlinear Schr\"{o}dinger equation for quartic soliton, dark soliton, cosine and elliptic periodic waves as weakly excited quantum waves propagating in He$^{4}$ films.  The phonon-roton dispersion equation for the elementary quantum excitations and cosine periodic solution are found in Sec. VI.
In this section, we have also presented the dimensionless form for nonlinear Schr\"{o}dinger equation describing the quantum dynamics in superfluid $^{4}$He films at low temperatures.
The numerical solutions based on this dimensionless nonlinear Schr\"{o}dinger equation are discussed in Sec. VII for traveling quantum waves. Finally, we give some conclusions in Sec. VIII.

\section{Nonlinear Schr\"{o}dinger equation for superfluid He$^{4}$ films}

We have formulated the quantum nonlinear Schr\"{o}dinger equation describing the superfluid He$^{4}$ in films at low temperatures when the quantum properties of fluid are significant: 
\begin{equation}
i\hbar\frac{\partial\psi}{\partial t}=\left(-\frac{\hbar^{2}}{2m}%
\nabla^{2}+V(|\psi|^{2})\right)\psi,  \label{1}
\end{equation}
\begin{equation}
V(|\psi|^{2})=G|\psi|^{2}-G|\psi_{0}|^{2}+\beta\nabla^{2}|\psi|^{2}+\sigma%
\nabla^{4}|\psi|^{2},  \label{2}
\end{equation}
where $m$ is the helium atomic mass. The wave function $\psi(x,y,t)$ in this nonlinear differential equation describes the superfluid helium in films at enough low temperatures. We use here the definitions: $\nabla=(\partial_{x},\partial_{y})$, $\nabla^{2}=\partial_{x}^{2}+\partial_{y}^{2}$ and $\nabla^{4}=%
\partial_{x}^{4}+2\partial_{x}^{2}\partial_{y}^{2}+\partial_{y}^{4}$. The complex wave function $\psi(x,y,t)$ can be presented in the following form, 
\begin{equation}
\psi(x,y,t)=U(x,y,t)\exp\left(\frac{i}{\hbar}\Theta(x,y,t)\right),
~~~~U(x,y,t)=\sqrt{\zeta(x,y,t)},  \label{3}
\end{equation}
where $\zeta(x,y,t)$ is the thickness of superfluid helium component in a film at point $(x,y)$ and time $t$. The superfluid in a ground state is uniform liquid at rest and 
the wave function of this ground state is $\psi=\psi_{0}=\mathrm{const}$.
Thus, we have $|\psi_{0}|^{2}=\zeta_{0}=\mathrm{const}$ where $\zeta_{0}$ is the thickness of superfluid helium component in a non excited (ground) state and the chemical potential is given as $G|\psi_{0}|^{2}=G\zeta_{0}$. The full thickness $d$ of helium film is $d(x,y,t)=\zeta_{n}+\zeta(x,y,t)$ where $\zeta_{n}=\mathrm{const}$ is a thickness of the normal component in helium film. Hence, the full thickness of helium film in the ground state is $d_{0}=\zeta_{n}+\zeta_{0}$ with $\zeta_{0}=\mathrm{const}$. The velocity $\mathbf{v}=(u,w)$ of superfluid helium in a film is 
\begin{equation}
\mathbf{v} =\frac{\hbar}{m}\nabla\Phi=\frac{1}{m}\nabla\Theta,  \label{4}
\end{equation}
where $\Phi(x,y,t)=\Theta(x,y,t)/\hbar$ is the phase of wave function $\psi(x,y,t)$. The
nonlinear Schr\"{o}dinger equation (\ref{1}) leads to the continuity equation: 
\begin{equation}
\frac{\partial\zeta}{\partial t}+\nabla\cdotp(\zeta\mathbf{v})=0,  \label{5}
\end{equation}
where $\zeta=|\psi(x,y,t)|^{2}$ and the velocity $\mathbf{v}(x,y,t)$ of superfluid helium is given by Eq. (\ref{4}).

Now we consider the classical limit for nonlinear Schr\"{o}dinger equation (\ref{1}). In this case we should assume the following constraint \cite{Sakurai}: 
\begin{equation}
\hbar|\nabla^{2}\Theta |\ll |\nabla\Theta|^{2}.  \label{6}
\end{equation}
The nonlinear Schr\"{o}dinger equation (\ref{1}) and condition given in Eq. (\ref{6}) yield the nonlinear partial differential equation: 
\begin{equation}
\frac{\partial\Theta}{\partial t} +\frac{1}{2m}|\nabla\Theta|^{2}+V(|%
\psi|^{2})=0.  \label{7}
\end{equation}
This is the Hamilton-Jacobi equation which first is found in classical mechanics. We note that $\Theta$ in classical mechanics is the Hamilton's principal function. Equation (\ref{7}) with Eqs. (\ref{2}) and (\ref{4}) yield
\begin{equation}
\frac{\partial\Theta}{\partial t} +\frac{m\mathbf{v}^{2}}{2}%
+G(\zeta-\zeta_{0}) +\beta\nabla^{2}\zeta+\sigma\nabla^{4}\zeta=0.  \label{8}
\end{equation}
The operator $m^{-1}\nabla$ applied to Eq. (\ref{8})
lead to the following equation for the velocity $\mathbf{v}$: 
\begin{equation}
\frac{\partial \mathbf{v}}{\partial t} +\frac{1}{2}\nabla (\mathbf{v}%
^{2})+G_{0}\nabla\zeta +\beta_{0}\nabla(\nabla^{2}\zeta)+\sigma_{0}
\nabla(\nabla^{4}\zeta)=0,  \label{9}
\end{equation}
with $G_{0}=G/m$, $\beta_{0}=\beta/m$ and $\sigma_{0}=\sigma/m$. Thus, the continuity Eq. (\ref{5}) and Eq. (\ref{9}) are the closed system of equations describing the classical fluid dynamics for the functions $\zeta(x,y,t)$ and $\mathbf{v}(x,y,t)$.
In the case when velocity $\mathbf{v}=(u,w)$ has only the component $u$ (with $w=0$) the system of Eqs. (\ref{5}) and (\ref{9}) for the functions $\zeta(x,t)$ and $u(x,t)$ has the form: 
\begin{equation}
\partial_{t}\zeta+ \partial_{x}(\zeta u)=0,  \label{10}
\end{equation}
\begin{equation}
\partial_{t}u+ u\partial_{x}u+G_{0}\partial_{x}\zeta
+\beta_{0}\partial_{x}^{3}\zeta +\sigma_{0}\partial_{x}^{5}\zeta=0.
\label{11}
\end{equation}

We have proved below that the system of Eqs. (\ref{10}) and (\ref{11}) describes the classical dynamics of gravity waves in incompressible fluid with appropriate definition of parameters $G_{0}=G/m$, $\beta_{0}=\beta/m$ and $\sigma_{0}=\sigma/m$.
One can define the surface deviation under the equilibrium level as $\eta=\zeta-\zeta _{0}$, then Eqs. (\ref{10}) and (\ref{11}) lead to the following linearized equations for the functions $\eta $ and $u$: 
\begin{equation}
\partial _{t}\eta +\zeta _{0}\partial _{x}u=0,~~~~\partial
_{t}u+G_{0}\partial _{x}\eta +\beta _{0}\partial _{x}^{3}\eta +\sigma
_{0}\partial _{x}^{5}\eta =0.  \label{12}
\end{equation}
Now we use the ansatz as
\begin{equation}
\eta =A(k)\exp [i(kx-\omega t)],~~~~u=B(k)\exp [i(kx-\omega t)],  \label{13}
\end{equation}
where $A(k)$ and $B(k)$ are real-valued quantities while $k$ and $\omega $ are the wave number and frequency respectively. Inserting this ansatz into Eq. (\ref{12}), one gets the following coupled equations for $A(k)$ and $B(k)$, 
\begin{equation}
-\omega A(k)+k\zeta_{0}B(k)=0,
\end{equation}
\begin{equation}
-\omega B(k)+kG_{0}A(k)-k^{3}\beta _{0}A(k)+k^{5}\sigma _{0}A(k)=0.
\label{15}
\end{equation}
This system of coupled equations has a nontrivial solution only when $k$ and $\omega $ obey the dispersion relation: 
\begin{equation}
\omega^{2}=\frac{G_{0}}{\zeta_{0}}(k\zeta_{0})^{2} -\frac{\beta_{0}}{%
\zeta_{0}^{3}}(k\zeta_{0})^{4} +\frac{\sigma_{0}}{\zeta_{0}^{5}}%
(k\zeta_{0})^{6}.  \label{16}
\end{equation}
The dispersion relation for the gravity waves in incompressible fluids is \cite{Whitham}: 
\begin{equation}
\omega^{2}=\left(gk+\frac{\gamma k^{3}}{\rho} \right)\tanh(k\zeta_{0}),
\label{17}
\end{equation}
where $g$ is the acceleration by gravity, $\gamma$ is the surface tension and $\rho$ is the mass density of helium liquid. Decomposition of Eq. (\ref{17}) can be written as 
\begin{equation}
\omega ^{2}=\frac{g}{\zeta _{0}}(k\zeta _{0})^{2}-\left( \frac{g}{3\zeta _{0}%
}-\frac{\gamma }{\rho \zeta _{0}^{3}}\right) (k\zeta _{0})^{4}+\left( \frac{%
2g}{15\zeta _{0}}-\frac{\gamma }{3\rho \zeta _{0}^{3}}\right) (k\zeta
_{0})^{6}+...~.  \label{18}
\end{equation}
Thus, Eqs. (\ref{16}) and (\ref{18}) lead to parameters $G_{0}$, $\beta_{0}$ and $\sigma_{0}$ as 
\begin{equation}
G_{0}=g,~~~~\beta_{0}=\frac{1}{3}g\zeta_{0}^{2} -\frac{\gamma}{\rho},
\label{19}
\end{equation}
\begin{equation}
\sigma_{0}=\frac{2g\zeta_{0}^{4}}{15} -\frac{\gamma\zeta_{0}^{2}}{3\rho},
\label{20}
\end{equation}
when the higher order terms in Eq. (\ref{18}) can be neglected.
It is shown in the Appendix A that the system of Eqs. (\ref{10}) and (\ref{11}) with the coefficients given in Eqs. (\ref{19}) and (\ref{20}) leads to Boussinesq equations for long gravity waves in incompressible fluids when we can neglect the last term in the left side of Eq. (\ref{11}). Thus, the derived system of Eqs. (\ref{10}) and (\ref{11}) is more general than Boussinesq equations. 

In conclusion, we have demonstrated that in the classical limit given by Eq. (\ref{6}) the found nonlinear Schr\"{o}dinger equation reduces to the generalized system of equations describing the propagation of long gravity waves in incompressible fluids. However, the description of quantum dynamics in superfluid helium films needs different definition for parameters $G$, $\beta$ and $\sigma$ in nonlinear Schr\"{o}dinger
equation (\ref{1}). Such definitions of the parameters for superfluid helium in films we preset in Sec. III and in more general form in Sec. VI respectively. 

\section{Dispersion equation for Nonlinear Schr\"{o}dinger equation}

In this section we derive the dispersion equation connected with generalized form of two-fluid hydrodynamics. This approach leads to particular definition for the parameters $G$, $\beta$ and $\sigma$ of nonlinear Schr\"{o}dinger equation (\ref{1}). We consider here the  wave function $\psi(x,t)$ depending on two variables $x$ and $t$. In this case the nonlinear Schr\"{o}dinger equation (\ref{1}) for superfluid He$^{4}$ films at low temperatures has the following form: 
\begin{equation}
i\hbar\frac{\partial\psi}{\partial t}=-\frac{\hbar^{2}}{2m}\frac{\partial
^{2}\psi}{\partial x^{2}}+ G\left(|\psi|^{2}-|\psi_{0}|^{2}\right)\psi+\beta%
\frac{\partial ^{2}|\psi|^{2}}{\partial x^{2}}\psi +\sigma\frac{\partial
^{4}|\psi|^{2}}{\partial x^{4}}\psi.  
\label{21}
\end{equation}
We present the wave function in the form $\psi(x,t)=U(x,t)\exp(i\Theta(x,t)/\hbar)$ with $U(x,t)=\sqrt{\zeta(x,t)}$ then the nonlinear Schr\"{o}dinger Eq. (\ref{21}) yields the system of equations: 
\begin{equation}
U_{t}=-\frac{1}{2m}\Theta_{xx}U-\frac{1}{m}\Theta_{x}U_{x},  \label{22}
\end{equation}
\begin{equation}
-\Theta_{t}=-\frac{\hbar^{2}}{2m}\frac{(\zeta^{1/2})_{xx}} {\zeta^{1/2}}+%
\frac{1}{2m} (\Theta_{x})^{2}+G(\zeta-\zeta_{0})+\beta\zeta_{xx}+\sigma
\zeta_{xxxx}.  \label{23}
\end{equation}
Equation (\ref{22}) can also be written as 
\begin{equation}
\zeta_{t}+\frac{1}{m}(\Theta_{x}\zeta)_{x}=0,  \label{24}
\end{equation}
which is the continuity equation.
Using the definition of helium velocity given in Eq. (\ref{4}) with $w=0$ we can write the velocity of helium as $u=m^{-1}\Theta_{x}$. Thus, Eq. (\ref{24}) yields the following standard form of continuity equation: 
\begin{equation}
\zeta_{t}+(u\zeta)_{x}=0.  \label{25}
\end{equation}

We can write the thickness of superfluid component of helium and the phase as $\zeta=\zeta_{0}+\eta$ and $\Theta=\Theta_{0}+\theta$ where $\eta$ is the deviation of surface for superfluid helium component from non exited depth $\zeta_{0}=\mathrm{const}$, $\Theta_{0}=\mathrm{const}$ is a phase for ground state of helium in the film and $\theta$ is deviation of the phase for superfluid component of helium. 
Linearization of the system of Eqs. (\ref{23}) and (\ref{24}) to small deviations $\eta$ and $\theta$ leads to the following linear equations: 
\begin{equation}
-\theta_{t}=-\frac{\hbar^{2}}{4m\zeta_{0}}\eta_{xx}+G\eta
+\beta\eta_{xx}+\sigma\eta_{xxxx},  \label{26}
\end{equation}
\begin{equation}
\eta_{t}=-\frac{\zeta_{0}}{m}\theta_{xx}.  \label{27}
\end{equation}
The system of Eqs. (\ref{26}) and (\ref{27}) has the following solution: 
\begin{equation}
\eta=a(k)\cos(kx-\omega t),~~~~ \theta=b(k)\sin(kx-\omega t),  \label{28}
\end{equation}
with the wave number $k$ and frequency $\omega$. The substitution of functions given in Eq. (\ref{28}) to Eqs. (\ref{26}) and (\ref{27}) yields the system of equations: 
\begin{equation}
b(k)\omega=\frac{\hbar^{2}a(k)k^{2}}{4m\zeta_{0}}+Ga(k) -\beta
a(k)k^{2}+\sigma a(k)k^{4},~~~~a(k)\omega=\frac{\zeta_{0}}{m}b(k)k^{2}.
\label{29}
\end{equation}
The solution of these linear equations leads to dispersion equation as 
\begin{equation}
\omega^{2}=\frac{G}{m\zeta_{0}}(k\zeta_{0})^{2} +\left(\frac{\hbar^{2}}{%
4m^{2}\zeta_{0}^{4}}-\frac{\beta} {m\zeta_{0}^{3}}\right)(k\zeta_{0})^{4} +%
\frac{\sigma}{m\zeta_{0}^{5}}(k\zeta_{0})^{6}.  \label{30}
\end{equation}

We also define the generalized dispersion equation for superfluid helium films which takes into account the roton excitations in helium superfluid: 
\begin{equation}
\omega^{2}=\left(\frac{c_{s}^{2}k}{\zeta_{0}} +\frac{q\gamma k^{3}}{\rho}
+qf_{r}\zeta_{0}^{4}k^{5}\right)\tanh(k\zeta_{0}).  \label{31}
\end{equation}
We note that in this equation, the first two terms in brackets follow from two-fluid hydrodynamics and the third term is connected with roton excitation. The third sound in the superfluid helium film is defined as $c_{s}=(d\omega/dk)_{k=0}$ and the full mass density is $\rho=\rho_{n}+\rho_{s}$ where $\rho_{n}$ and $\rho_{s}$ are the mass densities of normal and superfluid components respectively. The parameter $q$ is the superfluid fraction $q=\rho_{s}/\rho$ defined for suprfluid helium film in the ground state and $\gamma$ is the surface tension in helium film. The last term in brackets $qf_{r}\zeta_{0}^{4}k^{5}$ proportional to $k^{5}$ describes the roton effect for dispersion in superfluid helium film. The parameter $f_{r}$ is some characteristic acceleration and $F_{r}=mf_{r}$ is appropriate force connected with roton excitations in helium superfluid film. The generalized dispersion equation (\ref{31}) has the following decomposition with dimensionless parameter $k\zeta_{0}$: 
\begin{equation}
\omega^{2}=\frac{c_{s}^{2}}{\zeta_{0}^{2}}(k\zeta_{0})^{2} +\left(\frac{%
q\gamma}{\rho\zeta_{0}^{3}}-\frac{c_{s}^{2}} {3\zeta_{0}^{2}}%
\right)(k\zeta_{0})^{4} +\left(\frac{qf_{r}}{\zeta_{0}}+\frac{2c_{s}^{2}} {%
15\zeta_{0}^{2}}-\frac{q\gamma}{3\rho\zeta_{0}^{3}}\right)(k\zeta_{0})^{6}+...~.  \label{32}
\end{equation}
We can consider Eqs. (\ref{30}) and (\ref{32}) in the case with $k\zeta_{0}\ll 1$ which leads to explicit formulas for parameters $G$, $\beta$ and $\sigma$ as 
\begin{equation}
G=\frac{mc_{s}^{2}}{\zeta_{0}},~~~~ \beta=\frac{\hbar^{2}}{4m\zeta_{0}}-%
\frac{mq\gamma}{\rho} +\frac{m\zeta_{0}c_{s}^{2}}{3},  \label{33}
\end{equation}
\begin{equation}
\sigma=mqf_{r}\zeta_{0}^{4}+\frac{2}{15}mc_{s}^{2}\zeta_{0}^{3} -\frac{1}{%
3\rho}mq\gamma\zeta_{0}^{2}.  \label{34}
\end{equation}
We emphasize that more general form of dispersion equation with appropriate definition of parameters $G$, $\beta$ and $\sigma$ in nonlinear Schr\"{o}dinger equation (\ref{1}) for superfluid He$^{4}$ films at low temperatures is given in Sec. VI.

\section{Equation for weakly excited quantum waves in He$^{4}$ films}

In this section, we consider the traveling solitary and periodic waves described by Eq. (\ref{21}). In this case the functions $U(s)$, $\zeta(s)$ and $\Theta(s)$ depend only on variable $s=x-vt$ and Eqs. (\ref{6}) and (\ref{7}) have the form, 
\begin{equation}
v\Theta^{\prime}=-\alpha\frac{U^{\prime\prime}}{U}+\frac{1}{2m}
(\Theta^{\prime})^{2}+G(U^{2}-U_{0}^{2}) +\beta(U^{2})^{\prime\prime}
+\sigma(U^{2})^{\prime\prime\prime\prime},  \label{35}
\end{equation}
\begin{equation}
-v\zeta^{\prime}+\frac{1}{m}(\Theta^{\prime}\zeta)^{\prime}= 0.  \label{36}
\end{equation}
where $\alpha=\hbar^{2}/2m$ and the prime means the derivative to variable $s$. Integration of Eq. (\ref{36}) yields the equations: 
\begin{equation}
\Theta^{\prime}=mv+\frac{mC_{0}}{\zeta},~~~~\Theta=\Theta_{0}+mvs+mC_{0}\int
\zeta^{-1}ds,  \label{37}
\end{equation}
where $C_{0}$ and $\Theta_{0}$ are the integration constants. Equation (\ref{35}) with the function $\Theta^{\prime}$ given in Eq. (\ref{37}) leads to differential equation: 
\begin{equation}
EU^{2}=-\alpha UU^{\prime\prime}+\frac{mC_{0}^{2}}{2U^{2}} +GU^{4}+\beta
U^{2}(U^{2})^{\prime\prime} +\sigma U^{2}(U^{2})^{\prime\prime\prime\prime},
\label{38}
\end{equation}
where the energy parameter $E$ is given as 
\begin{equation}
E=\frac{1}{2}mv^{2}+G\zeta_{0}.  \label{39}
\end{equation}
We have also the following relation: 
\begin{equation}
UU^{\prime\prime}=\frac{1}{2}(U^{2})^{\prime\prime} -\frac{%
((U^{2})^{\prime})^{2}}{4U^{2}}=\frac{1}{2} \zeta^{\prime\prime}-\frac{%
(\zeta^{\prime})^2}{4\zeta}.  \label{40}
\end{equation}
Equation (\ref{38}) with relation (\ref{40}) leads to the following
nonlinear differential equation: 
\begin{equation}
\sigma\zeta^{\prime\prime\prime\prime}+\beta \zeta^{\prime\prime} -\frac{%
\alpha}{2\zeta}\zeta^{\prime\prime} +\frac{\alpha}{4\zeta^{2}}%
(\zeta^{\prime})^2 +\frac{mC_{0}^{2}}{2\zeta^{2}}+G(\zeta-\zeta_{0})=\frac{1%
}{2}mv^{2}.  \label{41}
\end{equation}
Subtraction of the term $mC_{0}^{2}/2\zeta_{0}^{2}$ from the left and right side of Eq. (\ref{41}) with $\zeta=\zeta_{0}+\eta$ yields the following equation: 
\begin{equation}
\sigma\eta^{\prime\prime\prime\prime} +\beta\eta^{\prime\prime} -\frac{\alpha%
}{2(\zeta_{0}+\eta)}\eta^{\prime\prime} +\frac{\alpha}{4(\zeta_{0}+\eta)^{2}}%
(\eta^{\prime})^2 -\frac{mC_{0}^{2}}{2\zeta_{0}^{2}}\frac{(2\zeta_{0}\eta
+\eta^{2})}{(\zeta_{0}+\eta)^{2}}+G\eta=\frac{1}{2}mv^{2} -\frac{mC_{0}^{2}}{%
2\zeta_{0}^{2}}.  \label{42}
\end{equation}

Equation (\ref{42}) can be considerably simplified for weakly excited
solitary waves when the following condition is satisfied: $%
\Lambda/\zeta_{0}\ll 1$ with $\Lambda=\max|\eta|$. Decomposition of
appropriate terms in Eq. (\ref{42}) to the second order of the fraction $\eta/\zeta_{0}$ and its derivatives leads to the nonlinear differential equation for weakly excited quantum waves in He$^{4}$ films (see the details in Appendix B): 
\begin{equation}
\sigma\eta^{\prime\prime\prime\prime}+\nu\eta^{\prime\prime}
+2\mu\eta\eta^{\prime\prime}+\mu(\eta^{\prime})^2+Q\eta^{2} +R\eta+F=0,
\label{43}
\end{equation}
with parameters $\nu$, $\mu$, $Q$, $R$ and $F$ as 
\begin{equation}
\nu=\beta-\frac{\alpha}{2\zeta_{0}},~~~~ \mu=\frac{\alpha}{4\zeta_{0}^{2}}%
,~~~~ Q=\frac{3mC_{0}^{2}}{2\zeta_{0}^{4}},  \label{44}
\end{equation}
\begin{equation}
R=G-\frac{mC_{0}^{2}}{\zeta_{0}^{3}},~~~~ F=\frac{mC_{0}^{2}}{2\zeta_{0}^{2}}%
-\frac{1}{2}mv^{2}.  \label{45}
\end{equation}

The boundary conditions for solitary waves described by differential
equations (\ref{41}) and (\ref{43}) have the following form: $%
\eta\rightarrow 0$ for $s\rightarrow \pm\infty$. These boundary conditions for solitary waves yield 
\begin{equation}
F=\frac{mC_{0}^{2}}{2\zeta_{0}^{2}}-\frac{1}{2}mv^{2}=0.  \label{46}
\end{equation}
Thus, Eqs. (\ref{41}) and (\ref{43}) lead to velocity $v$ for traveling solitary waves as 
\begin{equation}
v=\pm\frac{C_{0}}{\zeta_{0}},  \label{47}
\end{equation}
where $C_{0}$ is the integration constant introduced in Eq. (\ref{37}). The velocity $u$ of superfluid in helium film is given as $u=\Theta^{\prime}/m$ which yields by Eq. (\ref{37}) the following equation: 
\begin{equation}
u=v+\frac{C_{0}}{\zeta_{0}+\eta}.  \label{48}
\end{equation}

\section{Analytical solutions for traveling quantum waves}

In this section, the quartic soliton, dark soliton, cosine and elliptic periodic wave solutions are obtained analytically as weakly excited quantum waves propagating in He$^{4}$ films. First we consider the quartic soliton solution
of Eq. (\ref{43}) which can be written as 
\begin{equation}
\eta(s)=A\,\mathrm{sech}^{2}(p(s-s_{0})),  \label{49}
\end{equation}
where $s_{0}=\mathrm{const}$, $A$ is the amplitude and $p^{-1}$ is the width
of quartic soliton. Nonlinear differential Eq. (\ref{43}) yields three equations for parameters $A$, $p$ and $Q$ as 
\begin{equation}
A=\frac{15\sigma Q}{8\mu^{2}}-\frac{3\nu}{2\mu},  \label{50}
\end{equation}
\begin{equation}
p^{2}=\frac{Q}{4\mu}-\frac{\nu}{5\sigma},  \label{51}
\end{equation}
\begin{equation}
16\sigma p^{4}+4\nu p^{2}+R=0,  \label{52}
\end{equation}
where $R=G-(2/3)\zeta_{0}Q$. Equations (\ref{51}) and (\ref{52}) yield the equation for parameter $Q$ and hence for the integration constant $C_{0}$ because $C_{0}^{2}=2\zeta_{0}^{4}Q/3m$. Thus, we have the following equations for parameter $Q$ and velocity $v$ given by Eq. (\ref{47}): 
\begin{equation}
\frac{\sigma}{\mu^{2}}Q^{2}-\left(\frac{3\nu}{5\mu} +\frac{2\zeta_{0}}{3}%
\right)Q +G-\frac{4\nu^{2}}{25\sigma}=0,~~~~ v=\pm\zeta_{0}\sqrt{\frac{2Q}{3m%
}}.  \label{53}
\end{equation}

\begin{figure}[h]
\includegraphics[width=1\textwidth]{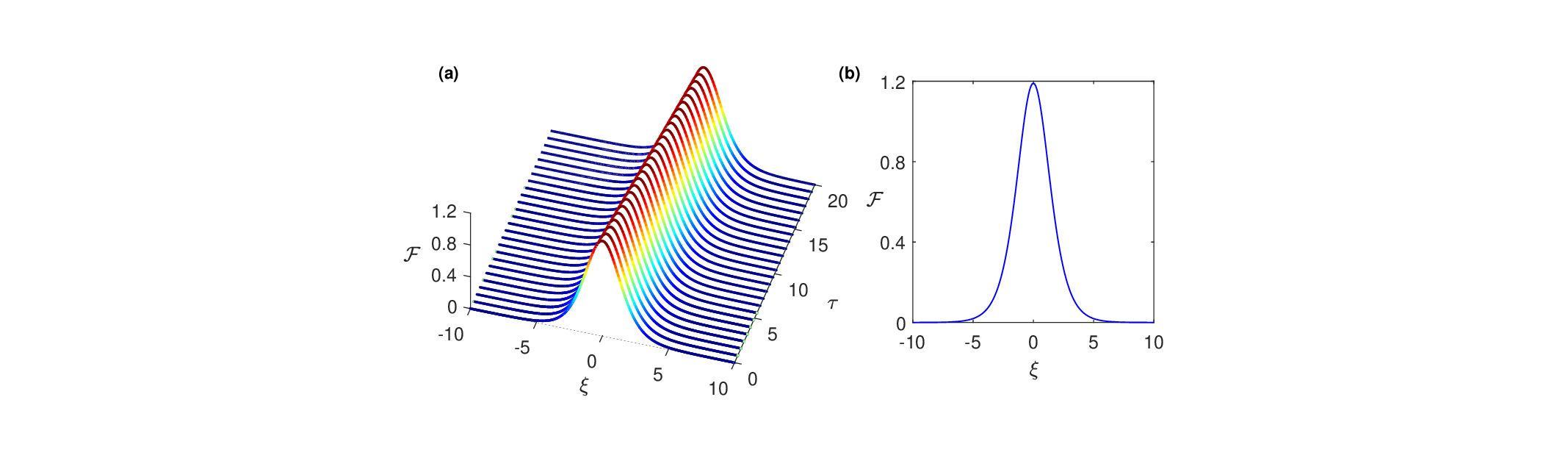}
\caption{Numerical evolution of the quartic soliton in superfluid He$^{4}$ film within the framework of Eq. (\ref{88}) for $c_{s}=59.36~m/s$, $\Delta/k_{B}=5.22~K$ and $k_{0}=2~\mathring{A}^{-1}$.}
\label{FIG.1.}
\end{figure}
\begin{figure}[h]
\includegraphics[width=1\textwidth]{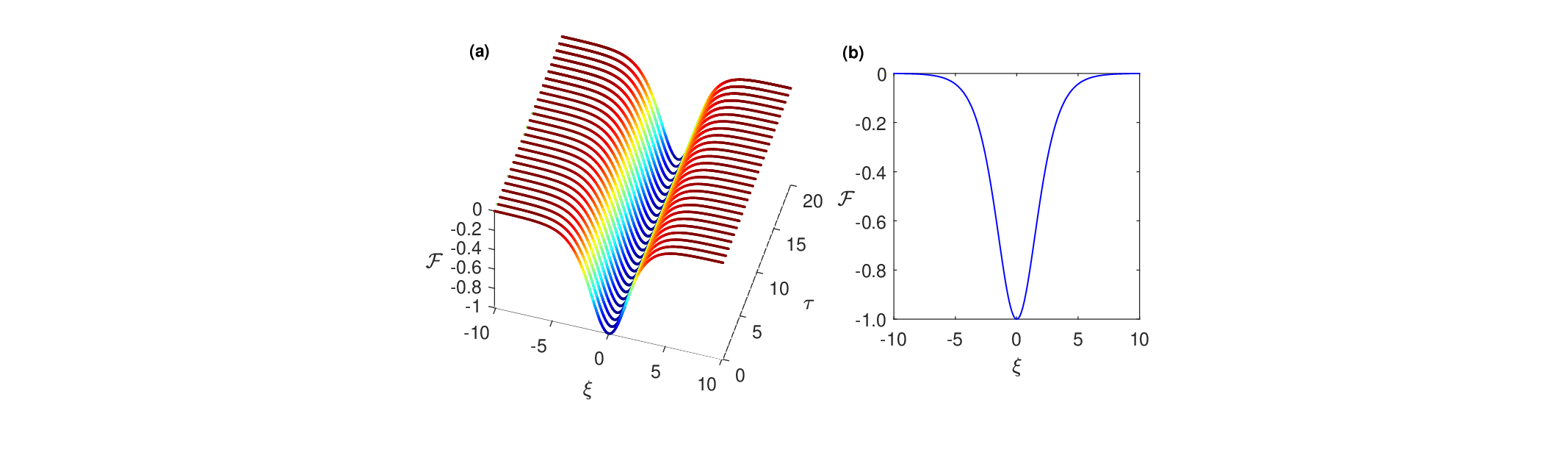}
\caption{Numerical evolution of the dark soliton in superfluid He$^{4}$ film within the framework of Eq. (88) for $c_{s}=59.36~m/s$, $\Delta/k_{B}=5.22~K$ and $k_{0}=2~\mathring{A}^{-1}$.}
\label{FIG.2.}
\end{figure}
\begin{figure}[h]
\includegraphics[width=1\textwidth]{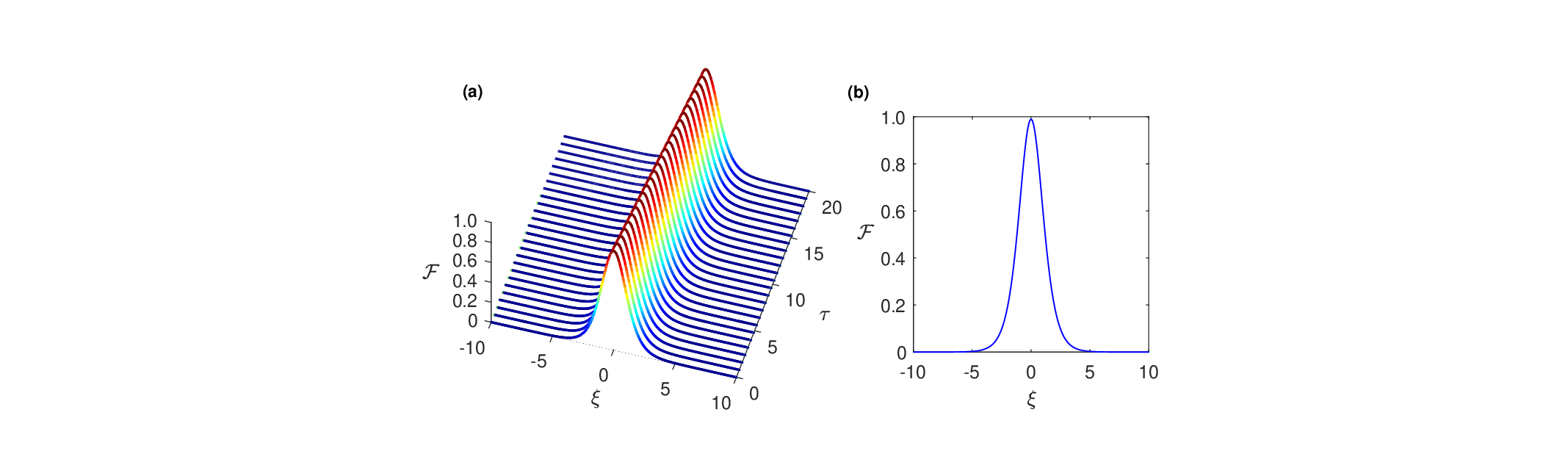}
\caption{Numerical evolution of the quartic soliton in superfluid He$^{4}$ film within the framework of Eq. (\ref{88}) for $c_{s}=63.4~m/s$, $\Delta/k_{B}=2.4~K$ and $k_{0}=0.8 ~\mathring{A}^{-1}$.}
\label{FIG.3.}
\end{figure}
\begin{figure}[h]
\includegraphics[width=1\textwidth]{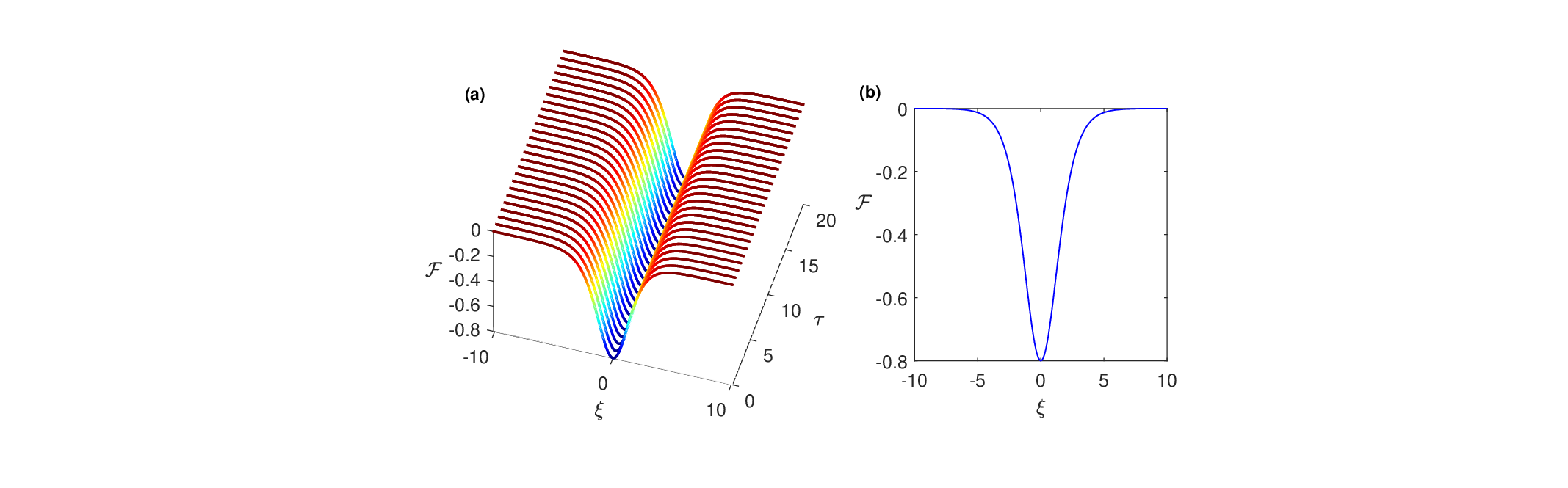}
\caption{Numerical evolution of the dark soliton in superfluid He$^{4}$ film within the framework of Eq. (\ref{88}) for $c_{s}=63.4~m/s$, $\Delta/k_{B}=2.4~K$ and $k_{0}=0.8 ~\mathring{A}^{-1}$.}
\label{FIG.4.}
\end{figure}

The substitution of parameter $Q$ from this quadratic equation to Eqs. (\ref{50}) and (\ref{51}) leads to explicit equations for the amplitude $A$ and inverse width $p$. Moreover, two necessary conditions follow for this quartic soliton solution: 
\begin{equation}
Q>0,~~~~\frac{Q}{4\mu}-\frac{\nu}{5\sigma}>0,  \label{54}
\end{equation}
where $Q$ is the appropriate positive solution of Eq. (\ref{53}). The
soliton solution presented in Eq. (\ref{49}) is correct when the condition $|\eta(s)|/\zeta_{0}\ll 1$ is satisfied because Eq. (\ref{43}) describes the weakly excited quantum waves in He$^{4}$ films. We emphasize that the amplitude $A$ can be positive or negative as well. In the case when amplitude $A$ given by Eq. (\ref{50}) is negative ($A<0$) we have the dark quartic soliton solution. Thus, we have the following solitary wave solution
of Eq. (\ref{21}): 
\begin{equation}
\psi(x,t)=\left[\zeta_{0}+A\,\mathrm{sech}^{2}(p(x-vt-s_{0})) \right]%
^{1/2}\exp\left(\frac{i}{\hbar} \Theta(x-vt)\right),  \label{55}
\end{equation}
where the function $\Theta(s)$ is given in Eq. (\ref{37}) with $s=x-vt$.

Equation (\ref{43}) for weakly excited quantum waves in He$^{4}$ films has the following periodic solution: 
\begin{equation}
\eta(s)=B\cos(q(s-s_{0})),  \label{56}
\end{equation}
where $s_{0}=\mathrm{const}$ and $B$ is the amplitude of periodic quantum waves. In this case Eq. (\ref{43}) yields three equations for parameters $B$, $q$ and $Q$ as 
\begin{equation}
\mu q^{2}B^{2}+\frac{mC_{0}^{2}}{2\zeta_{0}^{2}} -\frac{1}{2}mv^{2}=0,
\label{57}
\end{equation}
\begin{equation}
q^{2}=\frac{Q}{3\mu},  \label{58}
\end{equation}
\begin{equation}
\sigma q^{4}-\nu q^{2}+G-\frac{2}{3}\zeta_{0}Q=0.  \label{59}
\end{equation}
Thus, the parameter $Q$ and amplitude $B$ are given by the following
equations: 
\begin{equation}
\frac{\sigma}{3\mu^{2}}Q^{2}-\left(\frac{\nu}{\mu} +2\zeta_{0}\right)Q+3G=0,
\label{60}
\end{equation}
\begin{equation}
B=\pm\zeta_{0}\sqrt{\frac{v^{2}}{v_{0}^{2}}-1},~~~~ v_{0}^{2}=\frac{C_{0}^{2}%
}{\zeta_{0}^{2}}=\frac{2Q\zeta_{0}^{2}}{3m},  \label{61}
\end{equation}
and parameter $q$ is defined by Eqs. (\ref{58}) and (\ref{60}). We note that the velocity $v$ is a free parameter for periodic solution given in Eq. (\ref{56}). We note that for weakly excited periodic waves the following
condition should be satisfied: $|B|/\zeta_{0}\ll 1$ which yields the
condition for free parameter $v$ as $\sqrt{(v^{2}-v_{0}^{2})/v_{0}^{2}}\ll 1$. Hence, the periodic solution given in Eq. (\ref{56}) exists when the following conditions are satisfied: $Q>0$ where $Q$ is the appropriate
solution of Eq. (\ref{60}) and $v^{2}>v_{0}^{2}$ with 
$\sqrt{(v^{2}-v_{0}^{2})/v_{0}^{2}}\ll 1$. Thus, we have the following periodic wave solution of Eq. (\ref{21}): 
\begin{equation}
\psi(x,t)=\left[\zeta_{0}+B\cos(q(x-vt-s_{0})) \right]^{1/2}\exp\left(\frac{i%
}{\hbar} \Theta(x-vt)\right),  \label{62}
\end{equation}
where the function $\Theta(s)$ is given in Eq. (\ref{37}) with $s=x-vt$.

We have found that Eq. (\ref{43}) for weakly excited quantum waves in He$^{4} $ films has the periodic elliptic solution as 
\begin{equation}
\eta(s)=D\,\mathrm{cn}^{2}(p(s-s_{0}),k),  \label{63}
\end{equation}
where $s_{0}=\mathrm{const}$, $D$ is the amplitude and $\mathrm{cn}(z,k)$ is the Jacobi elliptic function of modulus $k$ with $0<k<1$. Using nonlinear differential Eq. (\ref{43}) we have found four equations for parameters $D$, $p$, $Q$ and velocity $v$: 
\begin{equation}
8\sigma Dp^{4}(2k^{2}-1)(1-k^{2})+2\nu Dp^{2}(1-k^{2}) +\frac{1}{3}%
\zeta_{0}^{2}Q-\frac{1}{2}mv^{2}=0,  \label{64}
\end{equation}
\begin{equation}
8\sigma p^{4}(2-17k^{2}+17k^{4})+4\nu p^{2}(2k^{2}-1) +8\mu
Dp^{2}(1-k^{2})+R=0,  \label{65}
\end{equation}
\begin{equation}
-120\sigma p^{4}k^{2}(2k^{2}-1)-6\nu p^{2}k^{2} +12\mu Dp^{2}(2k^{2}-1)+DQ=0,
\label{66}
\end{equation}
\begin{equation}
p^{2}=\frac{2\mu D}{15\sigma k^{2}}.  \label{67}
\end{equation}
Equations (\ref{66}) and (\ref{67}) lead to parameters $D$ and $p^{2}$ as 
\begin{equation}
D=\frac{15\sigma k^{2}Q}{8\mu^{2}(2k^{2}-1)} -\frac{3\nu k^{2}}{%
2\mu(2k^{2}-1)},  \label{68}
\end{equation}
\begin{equation}
p^{2}=\frac{Q}{4\mu(2k^{2}-1)}-\frac{\nu}{5\sigma(2k^{2}-1)}.  \label{69}
\end{equation}
Equation (\ref{67}) leads to relation $Dp^{2}=15\sigma k^{2}p^{4}/2\mu$. Using this relation we can write Eq. (\ref{65}) in the following form, 
\begin{equation}
4\sigma p^{4}(4-19k^{2}+19k^{4})+4\nu p^{2}(2k^{2}-1)+ G-\frac{2}{3}%
\zeta_{0}Q=0.  \label{70}
\end{equation}
Using Eq. (\ref{69}) we can also rewrite Eq. (\ref{70}) as 
\begin{equation}
\frac{4\sigma(4-19k^{2}+19k^{4})}{(2k^{2}-1)^{2}} \left(\frac{Q}{4\mu} -%
\frac{\nu}{5\sigma}\right)^{2} +\frac{4\nu(2k^{2}-1)}{(2k^{2}-1)}\left(\frac{%
Q}{4\mu} -\frac{\nu}{5\sigma}\right)+G-\frac{2}{3}\zeta_{0}Q=0.  \label{71}
\end{equation}
This quadratic equation yields the parameter $Q$ and hence Eqs. (\ref{68}) and (\ref{69}) with this value of $Q$ define the amplitude $D$ and parameter $p^{2}$. Moreover, using these parameters $Q$, $D$ and $p^{2}$ we can find the velocity $v$ from Eq. (\ref{64}). The necessary conditions for this periodic elliptic solution are 
\begin{equation}
Q>0,~~~~\frac{Q}{4\mu(2k^{2}-1)}-\frac{\nu}{5\sigma(2k^{2}-1)}>0,  \label{72}
\end{equation}
where $Q$ is the appropriate positive solution of Eq. (\ref{71}). Thus, we have the following periodic elliptic solution of Eq. (\ref{21}): 
\begin{equation}
\psi(x,t)=\left[\zeta_{0}+D\,\mathrm{cn}^{2}(p(x-vt-s_{0}),k) \right]%
^{1/2}\exp\left(\frac{i}{\hbar} \Theta(x-vt)\right),  \label{73}
\end{equation}
where the function $\Theta(s)$ is given in Eq. (\ref{37}) with $s=x-vt$. This elliptic solution for modulus $k=1$ reduces to solitary wave given in Eq. (\ref{55}).

In conclusion, we present the quartic soliton or quartic dark solution solution given by Eq. (\ref{49}) in dimensionless form: 
\begin{equation}
Y(\xi)=\mathrm{sign}(A)\frac{1} {\mathrm{cosh}^{2}(\xi-\xi_{0})},  \label{74}
\end{equation}
where $Y(\xi)=\eta(s)/|A|$ is dimensionless deviation of helium surface from non excited depth $\zeta_{0}$, $\xi=p(x-vt)$ is dimensionless variable, $\xi_{0}=ps_{0}$ is a free shift of soliton and $\mathrm{sign}(A)=A/|A|=\pm 1$ where the amplitude $A$ is given in Eq. (\ref{50}).

\section{Phonon-roton elementary excitations in superfluid He$^{4}$ films}

In this section, we define the coefficients $G$, $\beta$ and $\sigma$ in nonlinear Schr\"{o}dinger equation (\ref{1}) using the wave number $k_{0}$ at roton minimum $\Delta$. The parameters $k_{0}$ and $\Delta$ are connected by relation $E(k_{0})=\Delta$ where $E(k)$ is the energy of elementary excitations in superfluid He$^{4}$ films. In this case Eq. (\ref{30}) can also be written as 
\begin{equation}
E(k)=(\Lambda_{1}k^{2}+\Lambda_{2}k^{4} +\Lambda_{3}k^{6})^{1/2},  \label{75}
\end{equation}
where $E(k)=\hbar\omega(k)$ and the coefficients $\Lambda_{n}$ are 
\begin{equation}
\Lambda_{1}=\frac{G\zeta_{0}\hbar^{2}}{m},~~~~\Lambda_{2} =\frac{\hbar^{4}}{%
4m^{2}}-\frac{\beta\zeta_{0}\hbar^{2}}{m},~~~~\Lambda_{3}=\frac{%
\sigma\zeta_{0}\hbar^{2}}{m}.  \label{76}
\end{equation}
We can also write Eq. (\ref{75}) in the form: 
\begin{equation}
\varepsilon^{2}(p)=\lambda_{1}p^{2}+\lambda_{2}p^{4} +\lambda_{3}p^{6},
\label{77}
\end{equation}
with $\varepsilon(p)=E(k)$ and $p=\hbar k$. Then we have 
$\lambda_{1}=\Lambda_{1}\hbar^{-2}$, $\lambda_{2}=\Lambda_{2}\hbar^{-4}$ and 
$\lambda_{3}=\Lambda_{3}\hbar^{-6}$. We have the following conditions for coefficients $\lambda_{n}$ as 
\begin{equation}
\left(\frac{d\varepsilon(p)}{dp}\right)_{p=0}=c_{s}, ~~~~ \left(\frac{%
d\varepsilon(p)}{dp}\right)_{p=p_{0}}=0,~~~~ \varepsilon(p_{0})=\Delta,
\label{78}
\end{equation}
with $p_{0}=\hbar k_{0}$. These conditions lead to the following equations: 
\begin{equation}
\sqrt{\lambda_{1}}=c_{s},~~~~\lambda_{1}+2\lambda_{2}p_{0}^{2}+3%
\lambda_{3}p_{0}^{4}=0,  \label{79}
\end{equation}
\begin{equation}
\lambda_{1}p_{0}^{2}+\lambda_{2}p_{0}^{4} +\lambda_{3}p_{0}^{6}=\Delta^{2}.
\label{80}
\end{equation}
Thus, we have the coefficients $\lambda_{n}$ as 
\begin{equation}
\lambda_{1}=c_{s}^{2},~~~~ \lambda_{2}=\frac{3\Delta^{2}}{p_{0}^{4}} -\frac{%
2c_{s}^{2}}{p_{0}^{2}},~~~~ \lambda_{3}=\frac{c_{s}^{2}}{p_{0}^{4}} -\frac{%
2\Delta^{2}}{p_{0}^{6}},  \label{81}
\end{equation}
where $\lambda_{3}>0$, $\lambda_{2}<0$ and $\lambda_{1}>0$. The coefficients $\Lambda_{n}$ are given as 
\begin{equation}
\Lambda_{1}=c_{s}^{2}\hbar^{2},~~~~ \Lambda_{2}=\frac{3\Delta^{2}}{k_{0}^{4}}
-\frac{2c_{s}^{2}\hbar^{2}}{k_{0}^{2}},~~~~\Lambda_{3}= \frac{%
c_{s}^{2}\hbar^{2}}{k_{0}^{4}} -\frac{2\Delta^{2}}{k_{0}^{6}}.  \label{82}
\end{equation}
Thus, the energy of elementary excitations $E(k)$ has the following explicit form: 
\begin{equation}
E(k)=\left[c_{s}^{2}\hbar^{2}k^{2}+\left(\frac{3\Delta^{2}} {k_{0}^{4}}-%
\frac{2c_{s}^{2}\hbar^{2}}{k_{0}^{2}}\right)k^{4} +\left(\frac{%
c_{s}^{2}\hbar^{2}}{k_{0}^{4}} -\frac{2\Delta^{2}}{k_{0}^{6}}\right)k^{6}%
\right]^{1/2}.  \label{83}
\end{equation}

\begin{figure}[h]
\includegraphics[width=1\textwidth]{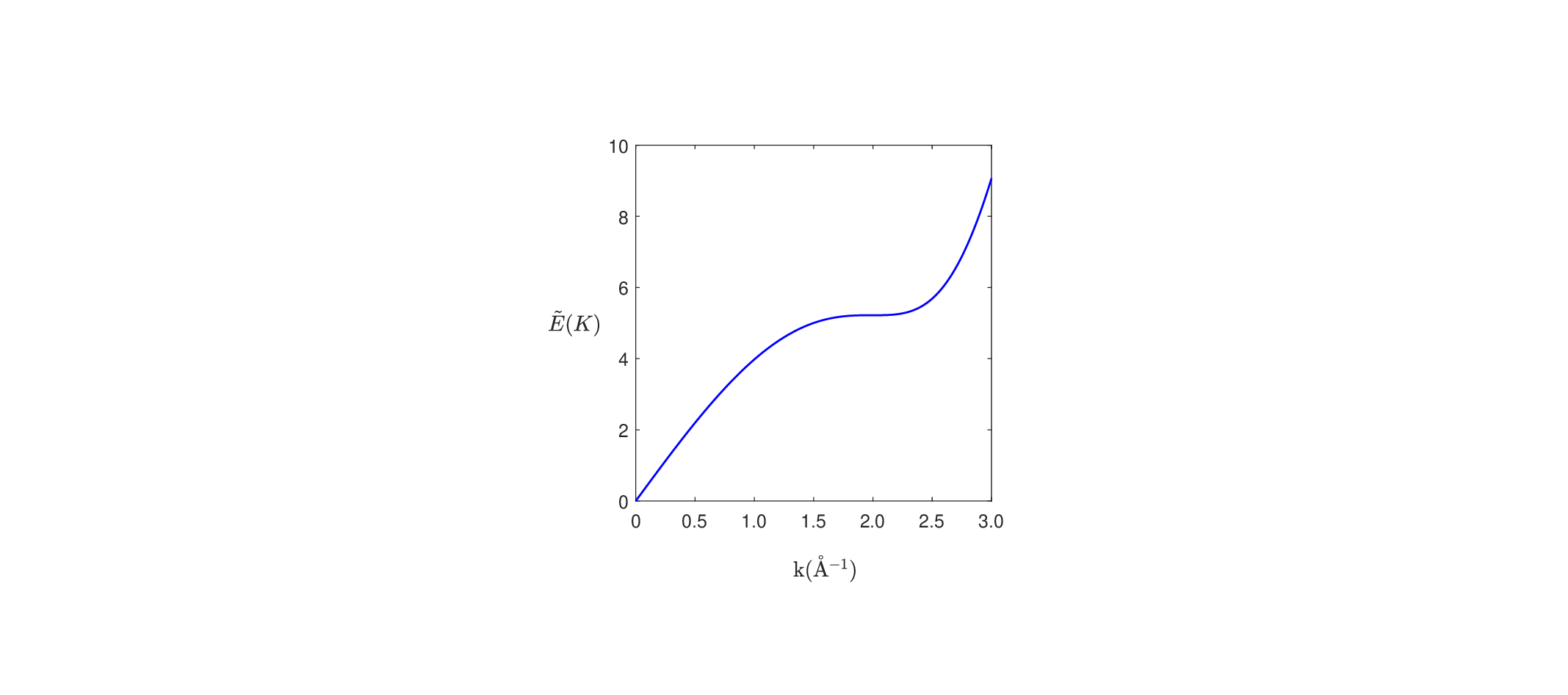}
\caption{Phonon-roton dispersion relation given by Eq. (\ref{98})
for $c_{s}=59.36~m/s$, $\Delta /k_{B}=5.22~K$ and $k_{0}=2~\mathring{A}^{-1}$. }
\label{FIG.5.}
\end{figure}
\begin{figure}[h]
\includegraphics[width=1\textwidth]{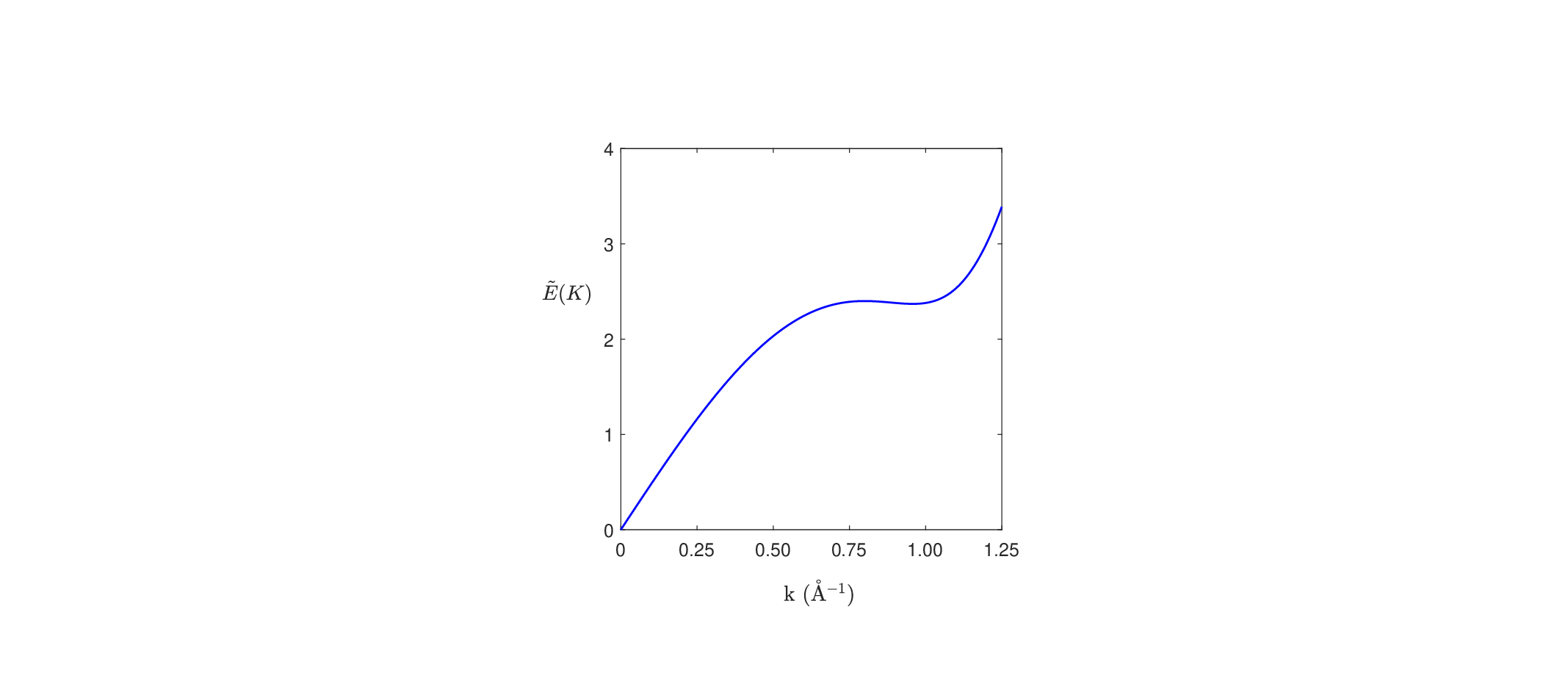}
\caption{Phonon-roton dispersion relation given by Eq. (\ref{98})
for $c_{s}=63.4~m/s$, $\Delta/k_{B}=2.4~K$ and $k_{0}=0.8~\mathring{A}^{-1}$.}
\label{FIG.6.}
\end{figure}
This equation yields the energy $E(k)$ at roton minimum $k=k_{0}$ as $E(k_{0})=\Delta$. Using the expansion of energy $E(k)$ in the vicinity of wave number $k=k_{0}$ we have found the equation (see Appendix C) which is correct for wave numbers $k$ near the roton minimum $k=k_{0}$: 
\begin{equation}
E(k)=\Delta+\frac{\hbar^{2}(k-k_{0})^{2}}{2m_{r}},  \label{84}
\end{equation}
where the effective mass for roton excitation is 
\begin{equation}
m_{r}=\frac{\hbar^{2}k_{0}^{2}\Delta}{4c_{s}^{2}
\hbar^{2}k_{0}^{2}-12\Delta^{2}}.  \label{85}
\end{equation}
Moreover, Eqs. (\ref{76}) and (\ref{82}) lead to the following expressions for parameters in nonlinear Schr\"{o}dinger equation (\ref{1}): 
\begin{equation}
G=\frac{mc_{s}^{2}}{\zeta_{0}},~~~~ \beta=\frac{\hbar^{2}}{4m\zeta_{0}} +%
\frac{2mc_{s}^{2}}{\zeta_{0}k_{0}^{2}} -\frac{3m\Delta^{2}}{%
\zeta_{0}\hbar^{2}k_{0}^{4}},  \label{86}
\end{equation}
\begin{equation}
\sigma=\frac{mc_{s}^{2}}{\zeta_{0}k_{0}^{4}} -\frac{2m\Delta^{2}}{%
\zeta_{0}\hbar^{2}k_{0}^{6}}.  \label{87}
\end{equation}

We use in our numerical simulations, discussed in Sec. VII, the  nonlinear Schr\"{o}dinger equation (\ref{21}) in dimensionless form with new variables $\tau=t/\delta$, $\xi=x/l$ and dimensionless wave function $\Psi(\xi,\tau)=\zeta_{0}^{-1/2}\psi(x,t)$. Here $\delta$ and $l$ are some arbitrary characteristic parameters of time and length respectively. In this case, the nonlinear Schr\"{o}dinger equation (\ref{21})
has the following dimensionless form: 
\begin{equation}
i\frac{\partial\Psi}{\partial \tau}
=-a_{0}\frac{\partial^{2}\Psi}{\partial \xi^{2}}
+a_{1}|\Psi|^{2}\Psi-a_{1}\Psi
+a_{2}\frac{\partial ^{2}|\Psi|^{2}}{\partial \xi^{2}}\Psi+a_{3}\frac{\partial^{4}|\Psi|^{2}}
{\partial \xi^{4}}\Psi.  
\label{88}
\end{equation}
We can use here the parameters $G$, $\beta$ and $\sigma$ given by Eqs. (\ref{86}) and (\ref{87}) which yield
the dimensionless coefficients $a_{n}$ as     
\begin{equation}
 a_{0}=\frac{\hbar \delta}{2ml^{2}},
 ~~~~a_{1}=\frac{mc_{s}^{2}\delta}{\hbar},
\label{89}
\end{equation}
\begin{equation}
a_{2}=\frac{\delta}{\hbar l^{2}}\left( \frac{\hbar^{2}}{4m}+\frac{2mc_{s}^{2}}{k_{0}^{2}} -\frac{3m\Delta^{2}}{\hbar^{2}k_{0}^{4}}\right),
\label{90}
\end{equation}
\begin{equation}
a_{3}=\frac{\delta}{\hbar l^{4}}\left(\frac{mc_{s}^{2}}{k_{0}^{4}} -\frac{2m\Delta^{2}}{\hbar^{2}k_{0}^{6}}\right).  
\label{91}
\end{equation}
We define the arbitrary parameters $\delta$ and $l$ as $\delta=\hbar(mc_{s}^{2})^{-1}$, $l=1/k_{0}$ which leads to coefficients $a_{n}$ in the following
form,
\begin{equation}
a_{0}=\frac{\hbar^{2}k_{0}^{2}}{2m^{2}c_{s}^{2}},
~~~~a_{1}=1,
\label{92}
\end{equation}
\begin{equation}
a_{2}=2+\frac{\hbar^{2}k_{0}^{2}}{4m^{2}c_{s}^{2}}
-\frac{3\Delta^{2}}{c_{s}^{2}\hbar^{2}k_{0}^{2}},
~~~~a_{3}=1-\frac{2\Delta^{2}}
{c_{s}^{2}\hbar^{2}k_{0}^{2}}.
\label{93}
\end{equation}
We emphasize that these coefficients $a_{n}$ are completely defined by parameters $c_{s}$, $\Delta$ and $k_{0}$.
We have found that the nonlinear Schr\"{o}dinger equation (\ref{88}) describing the superfluid He$^{4}$ films has the following periodic solution:
\begin{equation}
\Psi(\xi,\tau)=\sqrt{1+\cos(q_{0}(\xi-\xi_{0}))}
\exp(-i\Omega_{0}(\tau-\tau_{0})),
\label{94}
\end{equation}
where $\xi_{0}$ and $\tau_{0}$ are the arbitrary constant parameters and the dimensionless frequency $\Omega_{0}$ and parameter $q_{0}$ in this periodic solution are given as
\begin{equation}
\Omega_{0}=\frac{1}{4}a_{0}q_{0}^{2},
\label{95}
\end{equation}
\begin{equation}
a_{3}q_{0}^{4}-a_{2}q_{0}^{2}+a_{1}=0.	
\label{96}
\end{equation}
The nonlinear Schr\"{o}dinger equation (\ref{88}) with dimensionless coefficients $a_{n}$ given in Eqs. (\ref{92}) and (\ref{93}) is used in the Sec. VII for numerical simulations of localized quantum waves in the superfluid He$^{4}$ films at low temperatures.

\section{Numerical results for quantum waves and elementary excitations}

In this section, we study the wave evolution in the superfluid He$^{4}$ films by direct numerical simulations of Eq. (\ref{88}).
We consider below two different cases for experimental parameters $c_{s}$, $\Delta$ and $k_{0}$ presented in Ref. \cite{Rut} for superfluid He$^{4}$ films: case (1) $c_{s}=59.36~m/s$, $\Delta/k_{B}=5.22~K$, $k_{0}=2~\mathring{A}^{-1}$; case (2) $c_{s}=63.4~m/s$, $\Delta/k_{B}=2.4~K$, $k_{0}=0.8 ~\mathring{A}^{-1}$. 
Thus, we use in our numerical simulations the nonlinear Schr\"{o}dinger equation Eq. (\ref{88}) for two cases of parameters $a_{n}$ defined by Eq. (\ref{92}) and (\ref{93}).
We present in Figs. 1-4 for quartic solitons and dark solitons the function $\mathcal{F}(\xi,\tau)$ which is given as
\begin{equation}
\mathcal{F}(\xi,\tau)=\frac{1}{\zeta_{0}}(\zeta(x,t)
-\zeta_{0})=|\Psi(\xi,\tau)|^{2}-1.
\label{97}
\end{equation}

For solitons we have $\mathcal{F}(\xi,\tau)\geq 0$ because in this case $\zeta\geq \zeta_{0}$,
and for dark solitons $\mathcal{F}(\xi,\tau)\leq 0$
because for dark solitons $\zeta\leq \zeta_{0}$.
The propagation dynamics of localized quantum waves in the superfluid helium can be found by integrating the nonlinear Schr\"{o}dinger (\ref{88}) numerically. Here we utilize the split-step Fourier method for
solving Eq. (\ref{88}) and studying the dynamical evolution of the nonlinear waves
in the superfluid He$^{4}$ film. The input pulses are assumed to have a $\mathrm{sech}^{2}$ shape. Parameters used in the numerical simulation of Eq. (\ref{88}) are those given in the previously mentioned
two cases, which correspond to a realistic situation of wave propagation in
superfluid He$^{4}$ films. Figure 1 depicts the simulation result using the experimental parameters given in the case (1). From this figure, we can see that a quartic soliton can be readily generated in the system. Figure 2 exhibits the  profile in a uniform background that is a dark-type soliton excitation found numerically for parameters given in the case (1). In Figs. 3 and 4 we present a quartic soliton and a dark-type soliton for the parameters $c_{s}$, $\Delta/k_{B}$ and $k_{0}$ given in the case (2). One can see that the maximum amplitude of both quartic and dark solitons in Figs. 3 and 4 is decreased comparing to Figs. 1 and 2. Thus, the amplitude and width of propagating solitons are controlled through
the coefficients $a_{n}$ depending on physical parameters $c_{s}$, $\Delta/k_{B}$ and $k_{0}$. The numerical evolution results also show that the stable propagation is the main characteristic of the present soliton and dark-type soliton solutions.

We present in Figs. 5 and 6 the elementary excitations in superfluid He$^{4}$ films described by phonon-roton dispersion equation. We can write the phonon-roton dispersion equation given by Eq. (\ref{83}) in the following form: 
\begin{equation}
\tilde{E}(k)=(Ak^{2}+Bk^{4} +Ck^{6})^{1/2},  \label{98}
\end{equation}
where $\tilde{E}=E/k_{B}$ and $k=k(\mathring{A}^{-1})$. The coefficients $A$, $B$ and $C$ are given in this equation as 
\begin{equation}
A=c_{s}^{2}\hbar^{2}k_{B}^{-2}\cdot 10^{20},~~~~ B=3\tilde{\Delta}^{2}\tilde{%
k}_{0}^{-4} -2c_{s}^{2}\hbar^{2}k_{B}^{-2}\tilde{k}_{0}^{-2}\cdot 10^{20},
\label{99}
\end{equation}
\begin{equation}
C=c_{s}^{2}\hbar^{2}k_{B}^{-2}\tilde{k}_{0}^{-4}\cdot 10^{20}-2\tilde{\Delta}%
^{2}\tilde{k}_{0}^{-6},  \label{100}
\end{equation}
with $\tilde{\Delta}=\Delta/k_{B}$, and  $k_{0}=k_{0}(\mathring{A}^{-1})$.
Thus, Eqs. (\ref{99}) and (\ref{100}) yield the following coefficients for the cases (1) and (2): case (1) $A=20.5576~K^{2}\mathring{A}^{2}$, $B=-5.16972~K^{2}\mathring{A}^{4}$, $C=0.433336~K^{2}\mathring{A}^{6}$; case (2) $A=23.4511~K^{2}\mathring{A}^{2}$, $B=-31.0971~K^{2}\mathring{A}^{4}$, $C=13.3083~K^{2}\mathring{A}^{6}$.
Using these parameters $A$, $B$ and $C$
we exhibit in Figs. 5 and 6 the phonon-roton dispersion relation given in Eq. (\ref{98}) for cases (1) and (2) respectively.

\section{Conclusion}

In conclusion, we have presented a novel quantum nonlinear Schr\"{o}dinger equation describing the superfluid helium in film at enough low temperatures. It is shown that in classical limit the found nonlinear Schr\"{o}dinger equation for superfluid helium films reduces to a system of equations which are equivalent to Boussinesq equations for long gravity waves propagating in incimpressible fluids. It is also shown that the nonlinear Schr\"{o}dinger equation leads to phonon-roton dispersion relation for elementary excitations in superfluid He$^{4}$ films at enough low temperatures. We have found analytically the quartic solitons, dark solitons, periodic cosine and elliptic wave solutions for the weakly excited quantum waves in He$^{4}$ films. Our numerical simulations also demonstrate that the presented nonlinear Schr\"{o}dinger equation describes quartic and dark solitary waves in helium films. We emphasize  that these quantum solitary and periodic waves propagating on a continuous-wave background significantly differ from the analytical and numerical solutions obtained for the others model equations describing the dynamics of superfluid helium in film at enough low temperatures. We anticipate that obtained in this paper quantum waves and dispersion equation for the elementary phonon-roton excitations can find numerous  practical applications.

\appendix

\section{Equations for long gravity waves}

In this Appendix we show that the system of Eqs. (\ref{10}) and (\ref{11}) with the coefficients given in Eqs. (\ref{19}) and (\ref{20}) leads to Boussinesq and KdV equations describing the propagation of long gravity waves in incompressible fluids. We use here the standard notations accepted in the theory of
gravity waves: $\zeta\equiv h$ and $\zeta_{0}=d\equiv h_{0}$. We note that in the case when $(kh_{0})^{2}\ll 1$ one can neglect the last term $\sigma_{0}\partial_{x}^{5}\zeta$ in Eq. (\ref{11}). Equation (\ref{19}) with the new notations is 
\begin{equation}
G_{0}=g,~~~~\beta_{0}=\frac{1}{3}h_{0}c_{0}^{2} -\frac{\gamma}{\rho},
\label{a1}
\end{equation}
where $c_{0}=\sqrt{gh_{0}}$. Hence, the system of Eqs. (\ref{10}) and (\ref{11}) has the form: 
\begin{equation}
\partial_{t}h+ \partial_{x}(hu)=0,  \label{a2}
\end{equation}
\begin{equation}
\partial_{t}u+ u\partial_{x}u+g\partial_{x}h +\beta_{0}\partial_{x}^{3}h=0.
\label{a3}
\end{equation}
One can present the wave number as $k\simeq 1/l$ where $l$ is the characteristic length of the gravity wave. Hence, for long gravity waves we have the condition $(kh_{0})^{2}\ll 1$ or $\varepsilon^{2}=h_{0}^{2}/l^{2}\ll 1$. The last term in the left side of Eq. (\ref{a3}) has the order $\varepsilon^{2}\ll 1$. In this case one can use in the last term a lower approximation which is given by linear equation \cite{Whitham}: 
\begin{equation}
\partial_{t}^{2}h- c_{0}^{2}\partial_{x}^{2}h=0,  \label{a4}
\end{equation}
which yields the following relation $\partial_{x}%
\partial_{t}^{2}h=c_{0}^{2}\partial_{x}^{3}h$. The substitution of this approximation to Eq. (\ref{a3}) leads to the Boussinesq equations for the long gravity waves: 
\begin{equation}
\partial_{t}h+ \partial_{x}(hu)=0,  \label{a5}
\end{equation}
\begin{equation}
\partial_{t}u+ u\partial_{x}u+g\partial_{x}h
+\alpha_{0}\partial_{x}\partial_{t}^{2}h=0,  \label{a6}
\end{equation}
where $\alpha_{0}=\beta_{0}/c_{0}^{2}=h_{0}/3-\gamma/\rho c_{0}^{2}$. We emphasis that in Boussinesq equations the coefficient $\alpha_{0}$ has the form $\alpha_{0}=h_{0}/3$ because the effect of surface tension connected with parameter $\gamma$ was not considered earlier. Thus, we have shown that the system of Eqs. (\ref{10}) and (\ref{11}) is a generalization of Boussinesq equations. Moreover, the KdV equation follows from the Boussinesq equations when the additional parameter $\epsilon=(\max|h-h_{0}|)/h_{0}$ is small ($\epsilon\ll 1$). Hence, the Boussinesq and KdV equations follow from nonlinear Schr\"{o}dinger equation (\ref{1}) in the classical limit 
(see Eq. (\ref{6})) with additional conditions presented above. 

\section{Derivation of equation for weakly excited quantum waves}

In this Appendix we derive the nonlinear differential Eq. (\ref{43}) for weakly excited quantum waves in He$^{4}$ films. We can use here decomposition of the terms in Eq. (\ref{42}) to the second order of the fraction $\eta /\zeta _{0}$ and its derivatives: 
\begin{equation}
\frac{\eta ^{\prime \prime }}{\zeta _{0}+\eta }=\frac{\eta ^{\prime \prime }%
}{\zeta _{0}}-\frac{\eta \eta ^{\prime \prime }}{\zeta _{0}^{2}}+...~,
\label{b1}
\end{equation}%
\begin{equation}
\frac{(\eta ^{\prime })^{2}}{(\zeta _{0}+\eta )^{2}}=\frac{(\eta ^{\prime})^{2}}{\zeta _{0}^{2}}+...~,  \label{b2}
\end{equation}%
\begin{equation}
\frac{2\zeta _{0}\eta +\eta ^{2}}{(\zeta _{0}+\eta )^{2}}=\frac{2\eta }{\zeta _{0}}-\frac{3\eta ^{2}}{\zeta _{0}^{2}}+...~.  \label{b3}
\end{equation}%
The substitution of decomposition given in Eqs. (\ref{b1})-(\ref{b3}) to
Eq. (\ref{42}) leads to the following nonlinear differential equation for
weakly excited quantum waves in He$^{4}$ films: 
\begin{equation}
\sigma \eta ^{\prime \prime \prime \prime }+\nu \eta ^{\prime \prime }+2\mu\eta \eta ^{\prime \prime }+\mu (\eta ^{\prime })^{2}+Q\eta ^{2}+R\eta +F=0,
\label{b4}
\end{equation}%
where the parameters $\nu $, $\mu $, $Q$, $R$ and $F$ are given in Eqs. (\ref{44}) and (\ref{45}). Thus, this equation is derived for weakly excited quantum waves when the following condition is satisfied: $\Lambda /\zeta_{0}\ll 1$ with $\Lambda =\max |\eta |$.

\section{Dispersion equation near the roton minimum}

In this Appendix we derive the phonon-roton dispersion equation near the roton minimum of elementary excitation.
We can write the Taylor series for the energy of elementary excitations $E(k)$ at the roton minimum $k_{0}$ as 
\begin{equation}
E(k)=E(k_{0})+E^{\prime}(k_{0})(k-k_{0}) +\frac{1}{2}E^{\prime%
\prime}(k_{0})(k-k_{0})^{2}+...~,  \label{c1}
\end{equation}
where $E(k_{0})=\Delta$ and $E^{\prime}(k_{0})=0$. Equation (\ref{83}) yields the second derivative at $k=k_{0}$ as 
\begin{equation}
E^{\prime\prime}(k_{0})=\frac{4c_{s}^{2}\hbar^{2}}{\Delta} -\frac{12\Delta}{%
k_{0}^{2}}.  \label{c2}
\end{equation}
Thus, Eq. (\ref{c1}) in vicinity of the wave number $k=k_{0}$ can be written in the following form, 
\begin{equation}
E(k)=\Delta+\left(\frac{2c_{s}^{2}\hbar^{2}}{\Delta} -\frac{6\Delta}{%
k_{0}^{2}} \right)(k-k_{0})^{2}.  \label{c3}
\end{equation}
This equation can also be written as 
\begin{equation}
E(k)=\Delta+\frac{\hbar^{2}(k-k_{0})^{2}}{2m_{r}},  \label{c4}
\end{equation}
where the effective mass $m_{r}$ for roton excitation is 
\begin{equation}
m_{r}=\frac{\hbar^{2}k_{0}^{2}\Delta}{4c_{s}^{2}
\hbar^{2}k_{0}^{2}-12\Delta^{2}}.  \label{c5}
\end{equation}
We emphasize that the more general phonon-roton dispersion equation is given in Eq. (\ref{83}).

\end{document}